\documentstyle[12pt,epsfig,wrapfig,epsf]{article}

\renewcommand{\section}[1]{\vspace{6pt} \noindent\mbox{#1} \newline \noindent}
\renewcommand{\subsection}[1]{\vspace{6pt} \noindent\mbox{\underline{#1}} 
\newline \noindent}
\renewcommand{\subsubsection}[1]{\vspace{6pt} \noindent\mbox{\underline{#1}}
\noindent}

\newfont{\sansb}{cmssbx10}
\newfont{\sans}{cmss10}

\newcommand{\rb}[1]{\raisebox{1.5ex}[-1.5ex]{#1}}

\begin{document}
{\small OG 4.3.1 \vspace{-24pt}\\}     %The session code
{\center  \LARGE AGN STUDIES ABOVE 1.5 TeV WITH THE HEGRA 5\,m$^2$ CHERENKOV TELESCOPE
\vspace{6pt}\\}
D. Petry$^1$, S.M. Bradbury $^{1,}$\footnote{Present address: Dept. of Physics, University of Leeds, Leeds, UK},  A. Konopelko$^2$, D. Kranich$^1$, B.C. Raubenheimer$^3$
 and the\\ HEGRA collaboration \vspace{6pt}\\
{\it $^1$Max-Planck-Institut f\"ur Physik, F\"ohringer Ring 6, 80805 M\"unchen, Germany\\
$^2$Max-Planck-Institut f\"ur Kernphysik, 69029 Heidelberg, Germany\\
$^3$Space Research Unit, Dept. of Physics, 2520 Potchefstroom, South Africa 
 \vspace{-12pt}\\}
{\center ABSTRACT\\}
The HEGRA 5 m$^2$ air Cherenkov telescope (CT1) was used to search for gamma-ray emission above
 1.5\ TeV from a series of low redshift AGN (Mkn 421, Mkn 501, MS 0116+319, PKS 2209+236, NGC 315
and W Comae). Here we present results from a total of 383 hours of ON-source observations at zenith angles
up to 35$^\circ$ made between February 1996 and April 1997 showing positive detections of
Mkn 421 and Mkn 501 and upper limits on the other objects' emission.
More recent results from observations of Mkn 421 und Mkn 501 will be added at the conference.   

\setlength{\parindent}{1cm}
\section{INTRODUCTION}
Active Galactic Nuclei (AGN) of the BL Lac type are so far the only known extragalactic
sources of TeV gamma-rays\footnote{A preliminary detection of a third BL Lac object, 1ES2344+514, was presented
by the Whipple collaboration at the 4th Compton Symposium in April 1997.}.
After the discovery of Mkn 421 as a TeV gamma-ray source
by the Whipple observatory (Punch et al., 1992), it took another three years until
the discovery of Mkn 501 was made (Quinn et al., 1996), again
by the Whipple observatory. Both discoveries have in the meantime been confirmed by the HEGRA
experiment (Petry et al., 1996 and Bradbury et al., 1997, {\it = Papers I and II}) using two 
of the now six Cherenkov telescopes of the experiment. HEGRA
is now continuously monitoring both sources. The search for more sources of this type is continuing in the hope
to gain a larger sample such that general properties can be derived which is a prerequisite
of an understanding of the gamma-ray production mechanisms and the gamma-ray absorption due to the intergalactic
infrared background.
This paper summarizes the results obtained from AGN observations made with the HEGRA 5 m$^2$ Cherenkov
Telescope (CT1) during the past two years.

\section{THE TELESCOPE}
 CT1 is an equatorial mount $f/d = 1.7$ Cherenkov telescope with a high resolution Photomultiplier Camera consisting of
127 pixels in a 3.25$^\circ$ field of view (pixel diameter 0.25$^\circ$). The nominal energy threshold
of the telescope is 1.5 TeV although during the most recent observations we estimate the threshold
to be approx. 10 \% higher due to degradation of the mirror reflectivity.
The typical event rate is 0.7 Hz after application of quality cuts and the condition that at least 2 out of the 91 inner
pixels have to have a signal above 15 photoelectrons (``software trigger'').
Mount and mirrors of CT1 are described in Mirzoyan et al. (1994),
details of the camera and readout electronics are given by Rauterberg et al. (1995).
The telescope is part of the HEGRA cosmic ray detector system installed at 2200\,m a.s.l. on the
Canary island of La Palma (see OG 10.3.36).

\section{OBSERVED OBJECTS}
 Apart from the known sources Mkn 421 and Mkn 501, three of the  observed candiates
(MS 0116+319, PKS 2209+236 and W Comae) were chosen from a sample of 15 Blazars compiled by
Mannheim et al. (1996) which was available to us before publication.
All members of the sample are flat-spectrum radio sources with redshift $z <0.1$.
PKS 2209+236 and  W Comae have been detected by EGRET.
In addition, the Blazar NGC 315 was chosen for observation because of its extremely low redshift
of only 0.017. Another criterion for the selection was optimal observability from La Palma, 
i.e. a declination between 13$^\circ$ and 43$^\circ$.  

Both Mkn 421 and Mkn 501 have been shown to be highly variable sources which can
have short outbursts during which the emission can rise by factors of up to 50 (Gaidos et al., 1996,
Quinn et al., 1996). 
The non-detection of certain AGN at TeV energies is therefore only conclusive if it is over long observation times
or as part of a multi wavelength campaign which determines the overall state of the source.

\begin{table}[t] \centering
{\small
\begin{tabular}{|l|c|c|c|c|}
\hline
\multicolumn{5}{|c|}{HEGRA CT1 results from AGN observations in 1996/97 (part 1)} \\
\hline
Object Name                                                             & W Comae & PKS 2209+236 & NGC 315 & MS 0116+319 \\
\hline
number of nights of observations                                                    & 7       & 27     & 12 & 38 \\
\hline
first night (MJD)                 			                            & 50126   & 50275  & 50307 & 50347 \\
\hline
last night (MJD)                  			                            & 50141   & 50319  & 50321 & 50458 \\
\hline
total ON-time       (h)         			                            & 21.5   & 49.9    & 10.6 & 78.9 \\
\hline
maximum zenith angle ($^\circ$) 			                    	    & 30     & 30      & 18.7 & 30 \\
\hline
mean rate after filter (s$^{-1}$) 			                            & 0.83   & 0.64    & 0.64 & 0.72 \\
\hline
events after all cuts    				                            & 174    & 397     & 89   & 675 \\
\hline
expected background events    		& $167.9 \pm 8.6$ & $368.0 \pm 17.6$ & $ 79.9 \pm 7.1 $ & $ 632.7 \pm 30.6 $\\
\hline
excess events   			     & $6.1 \pm 15.7$ & $29.0 \pm 26.4 $ & $9.1 \pm 1.2$ & $42.3 \pm 40.4 $ \\ 
\hline
significance	($\sigma$)				                            & 0.39 & 1.1 & 0.78 & 1.05 \\
\hline
2 $\sigma$ U.L. on average rate   (h$^{-1}$)                      	            & 2.0 & 1.7 & 3.3 & 1.6 \\
\hline
2 $\sigma$ U.L. on average flux & & & & \\
 $F(E>1.5\,\mathrm{TeV})$ ($10^{-12}$\,cm$^{-2}$s$^{-1}$) & \rb{2.6} & \rb{2.2} & \rb{4.05} & \rb{2.1} \\
\hline
2 $\sigma$ U.L. on average (at $E$ = 1.5 TeV)   & & & & \\
of $\log_{10}(-E^2\mathrm{d}F/\mathrm{d}E/(\mathrm{erg\,cm}^{-2}\mathrm{s}^{-1})]$
	  \footnotemark[3] & \rb{-10.97} & \rb{-11.04} & \rb{-10.78} & \rb{-11.07} \\ 
\hline
2 $\sigma$ U.L. on 1-day ``flare''\footnotemark[4]  rate (h$^{-1}$)                 & 5.8 & 10.7 & 20.2 & 13.3 \\
\hline
2 $\sigma$ U.L. on 1-day ``flare''\footnotemark[4] flux & & & & \\
  $F(E>1.5\,\mathrm{TeV})$ ($10^{-12}$\,cm$^{-2}$s$^{-1}$) & \rb{7.7} & \rb{9.2} & \rb{25.0} & \rb{19.0} \\
\hline
expected number of chance occurences& & & & \\
 occurences of ``flare''\footnotemark[4]                   & \rb{0.57}  & \rb{1.1} & \rb{0.31} & \rb{0.39} \\
\hline
$\chi^2$ from constant fit to light curve                                           & 0.75 & 1.1 & 1.2 & 1.5 \\
\hline
\end{tabular}
}
\caption{\label{tab-ulims}
	Summary of the observations which resulted in upper limits. }
\end{table}

\section{OBSERVATIONS}
Tables \ref{tab-ulims} and \ref{tab-detect} show the observation times, the event statistics and the results
for all observations. 
For the object W Comae ($z$ = 0.102),  HEGRA took part in an international campaign in February
1996, preliminary results of which are described elsewhere (Maisack et al., 1997).
The other objects were observed independently.
The observations which led to the detection of Mkn 501 in 1996 are described in Paper II.
Here we present a reanalysis of this data applying new shower image parameter cuts.

\section{DATA ANALYSIS}
The data analysis including the background determination follows the method applied in Paper II
with the sole exception that
a new set of image parameter cuts was used. These so called {\it dynamical supercuts} have an improved
efficiency for gamma-showers and reach Q-factors of up to 11.5 (Kranich et al., 1997).

\begin{table}[t] \centering
{\small
\begin{tabular}{|l|c|c|c|}
\hline
\multicolumn{4}{|c|}{HEGRA CT1 results from AGN observations in 1996/97 (part 2)} \\
\hline
Object Name                                                         & Mkn 421 ('96) & Mkn 501 ('96) & Mkn 501 ('97) \\
\hline
number of nights of observations                                        & 17       &  70   & 22  \\
\hline
first night (MJD)                 			                & 50157   & 50169  & 50517  \\
\hline
last night (MJD)                  			                & 50222   & 50311  & 50553  \\
\hline
total ON-time       (h)         			                & 30.7  & 173.8    & 43.9  \\
\hline
maximum zenith angle ($^\circ$) 			                & 30     & 30  & 35  \\
\hline
mean rate after filter (s$^{-1}$) 			                & 0.74   & 0.74  & 0.65  \\
\hline
events after all cuts    				                & 311   & 1601 & 1002    \\
\hline
expected background events    		& $211.8 \pm 6.4$ & $1149.0 \pm 58.7 $ & $ 275.9 \pm 13.6 $ \\
\hline
excess events   			     & $99.2 \pm 20.1$ & $452.0 \pm 71.2 $ & $726.1 \pm 34.4 $ \\ 
\hline
significance	($\sigma$)				   & 4.7 & 6.4  & 21.1 \\
\hline
average excess rate   (h$^{-1}$)                      	            & $3.2 \pm 0.7$ & $2.6 \pm 0.4$  & $16.5 \pm 0.8$  \\
\hline
integral spectral index (preliminary)                           & 1.7\footnotemark[5] & $1.6\pm0.3$ & $1.8\pm0.2$ \\ 
\hline
 average flux $F(E>1.5\,\mathrm{TeV})$ ($10^{-12}$\,cm$^{-2}$s$^{-1}$) & $2.9 \pm 0.6$ & $2.3 \pm 0.4$ & $17.6 \pm 0.08$ \\
\hline
 systematic error on flux ($10^{-12}$\,cm$^{-2}$s$^{-1}$) & $+1.5-0.7$ & $+1.5-0.6$  & +7.5-4.1 \\
\hline
 average (at $E$ = 1.5 TeV) of & & & \\
$\log_{10}(-E^2\mathrm{d}F/\mathrm{d}E/(\mathrm{erg\,cm}^{-2}\mathrm{s}^{-1})]$ \footnotemark[3]
	& \rb{-10.93} & \rb{-11.03}  & \rb{-10.14}  \\ 
\hline
peak daily rate (h$^{-1}$)                 & $10.1 \pm 3.1$ & $8.0 \pm 2.0$ &   $53.2 \pm 5.1$ \\
\hline
MJD of peak                                & 50159.0 & 50251.0 & 50551.1 \\
\hline
peak daily flux $F(E>1.5\,\mathrm{TeV})$ ($10^{-12}$\,cm$^{-2}$s$^{-1}$) & $9.2 \pm 2.8$ & $7.2 \pm 1.8$  & $58.8 \pm 5.7$ \\
\hline
 systematic error on peak flux ($10^{-12}$\,cm$^{-2}$s$^{-1}$) & +3.9-2.1 & +3.1-1.7  & +25.2-13.5 \\
\hline
expected number of chance occurences of& & &  \\
observed flux peak assuming constant flux       & \rb{0.38}  & \rb{0.65} & \rb{$<6 \times 10^{-6}$} \\ 
\hline
$\chi^2$ from constant fit to light curve              & 0.85 & 1.3  & 6.9 \\
\hline
\end{tabular}
}
\caption{\label{tab-detect}
	Summary of the observations which resulted in detections. }
\end{table}

\footnotetext[3]{Calculated for the comparison with theoretical predictions assuming a differential spectral index $\alpha = 2.7$}
\footnotetext[4]{This is assuming the scenario of a single one-day flare occuring during the observation time.}
\footnotetext[5]{The integral spectral index 1.7 is assumed for the flux calculation; a measurement of the  index
will be presented at the conference. Preliminary investigations show Mkn 421 to have a steeper spectrum than
Mkn 501.}

The background for all observations was determined from a set of 142.8 h of OFF-source observations
at zenith angles up to 56$^\circ$ which were taken between Summer 1995 and Spring 1997. The method
relies on the fact that a pure set of background events is obtained when the image parameter ALPHA
is required to satisfy $20^\circ < \mathrm{ALPHA} < 80^\circ$. The event rate after this cut must be the same
within the statistical errors in the ON- and the OFF-source observation. In this
way a means to normalize the background is found. The statistical error of the normalisation is
taken into account in the calculation of the significance of the excess, and in order to account
for possible changes of the shape of the ALPHA-distribution, the normalisation is done seperately in several
zenith angle bins.

The method was tested by subdividing the OFF-dataset arbitrarily into 6 sets and calculating
the excess of each set over the background as determined from the remaining 5 sets: Although
the zenith angle distributions of the 6 sets were very different, no systematic excess could be found,
the sigificances from the 6 tries being -0.66, -0.83, +0.64, +1.0, -0.63 and +0.022 (average $-0.076\pm 0.219$).

The upper limits were calculated using the method described by Helene (1983) and adding a 50\% systematic error due
to uncertainties of the energy calibration and the source spectrum.

\begin{figure} \centering \leavevmode
\epsfxsize11.5cm
\epsffile{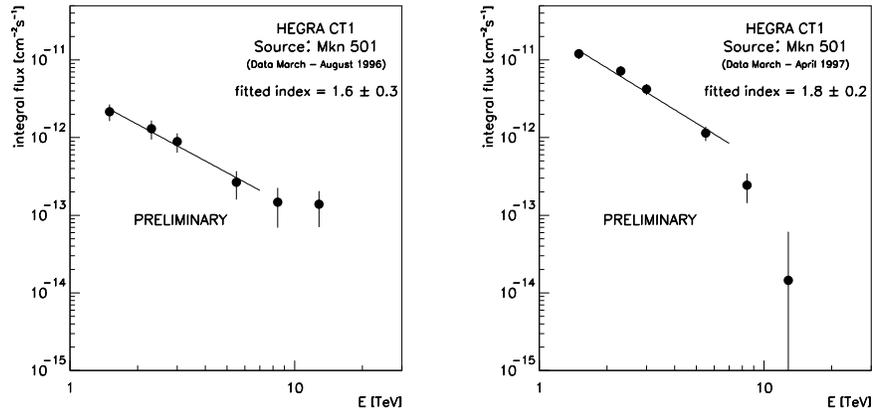}
\vspace{-1.2cm}
\caption{\label{fig}
	The preliminary spectra of Mkn 501 in 1996 (left) and 1997 (right). 
	The systematic errors on the two uppermost points are still under investigation.	        
}
\vspace{-0.3cm}
\end{figure}

\section{RESULTS}
The results are summarised in tables \ref{tab-ulims} and \ref{tab-detect} and are partially
preliminary. We obtain upper limits to
the emission of  W Comae, PKS 2209+236, NGC 315 and MS 0116+319. The upper limits on the average flux
are above the predicted flux values from Mannheim et al. (1996) coming close to the predictions in the
case of W Comae and MS 0116+319. For PKS 2209+236, the upper limit only agrees with the predictions
if external absorption is assumed. The limit on the flux of a single one-day-flare during the observation time
is between 1.0 and 3.0 Crab for all of the sources, but all lightcurves agree well with constant zero emission.

Positive detections are reported
for Mkn 421 and Mkn 501 in the 1996 observing period and for Mkn 501 in the 1997 observing period. The latter
is also separately analysed in OG 4.3.2 and from observations with the HEGRA telescope system in OG 10.3.19.

In 1996, both Mkn 421 and Mkn 501 were at quiescent levels of about 
0.3 Crab showing no significant flare behaviour.
However, in this year's observing period (1997, which is still ongoing
at the time of the preparation of this paper) Mkn 501 was an extremely bright object at
TeV energies reaching apparent luminosities of several times that of the Crab Nebula.
Figure \ref{fig} shows preliminary spectra of Mkn 501 in the 1996 and the 1997 observing
period which seems to indicate that the spectrum is steeper in the high state.

\section{ACKNOWLEDGEMENTS}
The HEGRA collaboration would like to thank the Instituto Astrofisico de Canarias for the use
of the HEGRA site and its facilities. This work has been supported by the BMBF, the DFG and the CICYT.

\section{REFERENCES}
\setlength{\parindent}{-5mm}
\begin{list}{}{\topsep 0pt \partopsep 0pt \itemsep 0pt \leftmargin 5mm
\parsep 0pt \itemindent -5mm}
\vspace{-15pt}

\item	Bradbury, S.M., Deckers, T., Petry, D., et al., A\&{}A, 320, L5, (1997).
\item	Gaidos, J.A., et al., Nature, 383, 319, (1996).
\item   Helene, O., NIM, 212, 319, (1983).
\item	Maisack, K., et al., Proc. ``Girona Conference on Blazars, Black Holes and Jets 1996'', to be published
in Astrophysics	and Space Science, Kluwer Acad. Publ.
\item   Kranich, D. \& Petry, D., to be presented at the Workshop ``Towards a major atmospheric Cherenkov
detector V'', Kruger National Park, (1997). 
\item	Mannheim, K., Westerhoff, S., Meyer, H. \& Fink, H.-H., A\&A, 315, 77, (1996). 
\item	Mirzoyan, R., Kankanian, R., Sawallisch, P., et al., NIM A, 351, 513, (1994).
\item	Petry, D., Bradbury, S.M., Konopelko, A., et al., A\&{}A, 311, L13, (1996).
\item	Punch, M., et al., Nature, 160, 477 (1992).
\item	Quinn, J., et al., ApJ, 456, L83, (1996).
\item	Rauterberg, G., et al., Proc. 24th Int. Cosmic Ray Conf., 3, 460, (1995).

\end{list}

\end{document}